# Photoinduced charge injection from shallow point defects in diamond into water

Kang Xu[1], Daniela Pagliero[1], Gabriel I. Lopez Morales[1], Johannes Flick[1,2,3], Abraham Wolcott[1,4], and Carlos A. Meriles[1,2,†]

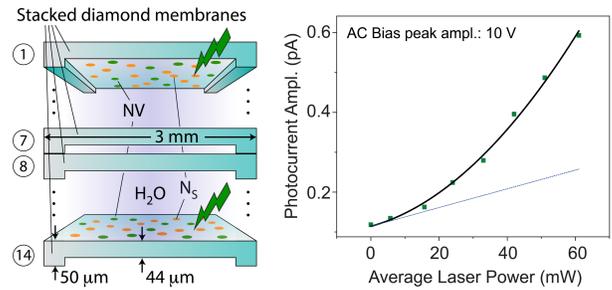

**Abstract**: Thanks to its low or negative surface electron affinity and chemical inertness, diamond is attracting broad attention as a source material of solvated electrons produced by optical excitation of the solid-liquid interface. Unfortunately, its wide bandgap typically imposes the use of wavelengths in the ultra-violet range, hence complicating practical applications. Here we probe the photocurrent response of water surrounded by single-crystal diamond surfaces engineered to host shallow nitrogen-vacancy (NV) centers. We observe clear signatures of diamond-induced photocurrent generation throughout the visible range and for wavelengths reaching up to 594 nm. Experiments as a function of laser power suggest that NV centers and other co-existing defects — likely in the form of surface traps — contribute to carrier injection, though we find that NVs dominate the system response in the limit of high illumination intensities. Given our growing understanding of near-surface NV centers and adjacent point defects, these results open new perspectives in the application of diamond-liquid interfaces to photo-carrier-initiated chemical and spin processes in fluids.

## Introduction

Injection, migration, and collection of photogenerated charges across solid-fluid interfaces are central to applications in electrochemistry[1,2], photocatalysis[3], and biology[4]. From a growing palette of source materials[5,6], diamond is drawing broad interest because it is chemically inert and its low or negative surface electron affinity facilitates the formation of solvated charges[7,8]. Electron injection from diamond into aqueous environments is attracting special attention given the electron's capacity to trigger reactions otherwise difficult to induce catalytically; examples include the reduction of dissolved $N_2$ to $NH_3$ or $CO_2$ into $CO$[7,9,10]. Carrier injection from diamond should also prove relevant to the study of confined fluidic environments, for example, to help understand ion transport in bio-membranes[11], or investigate memristic responses in nanofluidic channels[12]. Along related lines, the ability to optically inject carriers previously spin polarized via optical and/or microwave manipulation, could open new routes toward dynamic nuclear polarization in fluids[13,14], or in the exploration of spin-selective transport across liquid-solid interfaces[15]. We note that although other materials with favorable electron affinity are known[16,17], most react with water, which further singles out diamond as a preferred platform.

In the absence of mid-bandgap states, free carrier excitation requires the use of ultraviolet (UV) illumination with wavelength of 227 nm or shorter. Given the impracticalities of UV light, different strategies are being explored to reduce the photon energy to the visible range; these include functionalization with dye molecules[18], plasmonic coupling via embedded metal nanoparticles[19], surface nano-structuring[20], and the use of dopants such as boron, nitrogen, or phosphorous[21]. Experiments using ultrafast transient absorption spectroscopy demonstrated the injection of solvated electrons into water from surface-terminated detonation nanodiamond under 400 nm laser excitation[22]. Additionally, emission spectroscopy experiments on thin polycrystalline diamond films subject to light excitation of variable wavelength have shown electron injection into vacuum throughout the near-UV and visible range, from 340 to 550 nm[23].

Here, we implement photo-current measurements to probe carrier injection into ultrapure water from single-crystal, oxygen-terminated diamond engineered to feature optimal nitrogen-vacancy (NV) concentration from annealing-induced conversion of shallow-implanted nitrogen. Already exploited as nanoscale sensors[24-26], NV centers promise opportunities as a source of solvated carriers because visible light induces a cycle of charge state conversion[27], from negatively charged ($NV^-$) to neutral ($NV^0$) and back, respectively resulting in the generation of free electrons and holes[28,29]. With the help of a microfluidics chip tailored to yield a large contact area between water and single crystal diamond, we demonstrate steady-state photocurrent generation in water under visible light, which we attribute to a subtle interplay between contributions from NV centers — dominant at higher laser intensities — and photo-activated surface traps.

---

[1]Department. of Physics, CUNY-City College of New York, New York, NY 10031, USA. [2]CUNY-The Graduate Center, New York, NY 10016, USA. [3]Center for Computational Quantum Physics, Flatiron Institute, New York, NY 10010, USA. [4]Department of Chemistry, San José State University, San José, CA 95192, USA. [†]E-mail: cmeriles@ccny.cuny.edu

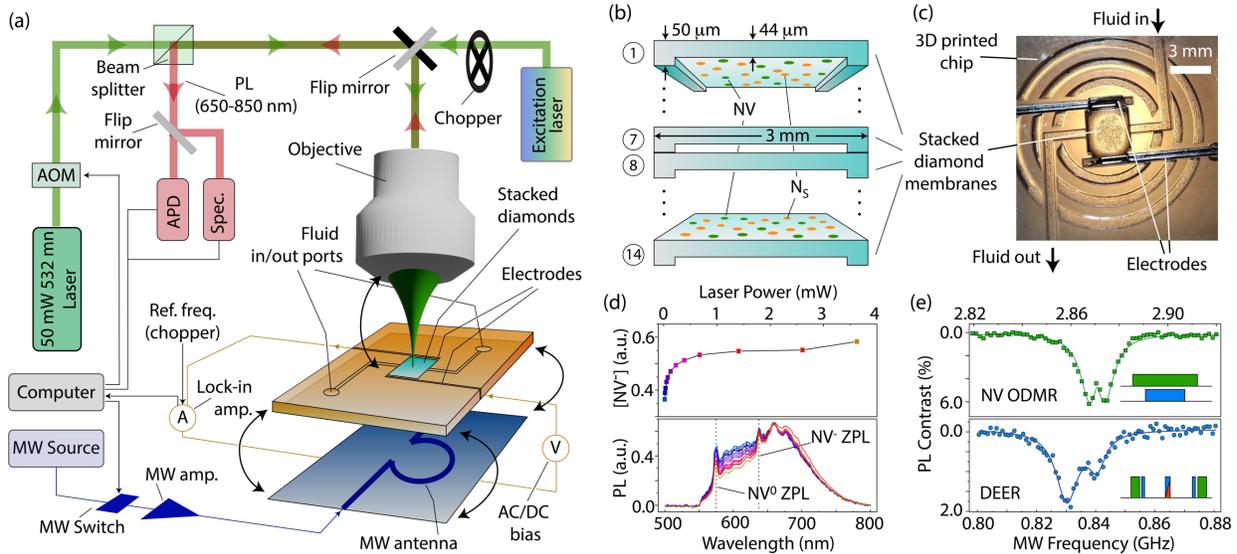

**Figure 1: Optofluidic setup.** (a) Schematic of the experimental setup. We use a microfluidic device to circulate a fluid through the 6-μm-tall spacings in a stack of fourteen 44-μm-thick, 3×3 mm² diamond membranes hosting 5-nm-deep NVs on either side. The system can be configured to operate as a confocal microscope or for photocurrent measurements. In this latter case, we apply a bias AC or DC voltage with the aid of two lateral gold-coated electrodes in contact with the fluid, and measure the photo-current using a lock-in amplifier. (b) Side view schematics of the diamond membrane stack. (c) Optical image of the microfluidic chip; adhesive in the circular trenches around the diamond chamber holds a thin glass cover slip serving as the upper seal. (d) Fluorescence spectroscopy of a representative diamond membrane for optical power growing from 12 μW (blue) to 3.6 mW (orange). The upper plot shows the fractional NV⁻ population as extracted from fits to the optical spectra. (e) Continuous-wave ODMR of the NVs in one of the membranes and NV-detected DEER spectrum of coexisting paramagnetic impurities (upper and lower plots, respectively). Solid traces indicate fits of two Lorentzian curves. Green blocks indicate 532 nm laser pulses while blue and red blocks respectively represent MW pulses resonant with the NVs and coexisting paramagnetic centers, see Methods. In (d) and (e), we configure the microscope in confocal mode, i.e., we excite and collect luminescence from a ~1 μm² area in one of the NV layers of a single, representative diamond membrane. AOM: Acousto-optic modulator. APD: Avalanche photodetector. Amp: Amplifier. PL: Photoluminescence. Spec.: PL spectrometer. ZPL: Zero-phonon line. Ref. Freq.: Reference frequency from chopper.

**Experimental approach and diamond characteristics**

Figure 1a lays out our setup: The key piece in our experiments is a microfluidic device designed to host a stack of fourteen 44-μm-thick membranes made from electronic grade, single-crystal diamond. $^{14}$N implantation on both sides of each membrane followed by thermal annealing yield ~5-nm-deep NV layers with concentration of $3 \times 10^{11}$ cm$^{-2}$. From the implantation conditions, we estimate a nitrogen content of $1 \times 10^{13}$ cm$^{-2}$, which corresponds to an NV formation efficiency of 3%, characteristic for near-surface centers[30] (see Methods). Exposure to acid mixtures prior to usage ensures predominantly oxygen-terminated surfaces[31] though the coverage is likely chemically and spatially heterogeneous with varying content of carbonyl and hydroxyl groups; remarkably, the surface electron affinity may remain negative due to persistent C–H bonds[8,31-33]. Throughout all charge injection experiments, we slowly flow water through the 6-μm openings between adjacent membranes as well as the 30 μm gap space between the electrodes and diamonds (Figs. 1b and 1c). We use collimated optical excitation of variable wavelength to homogeneously illuminate a 2-mm-diameter area of the membranes, and measure the resulting photo-current with the help of a lock-in amplifier referenced to an optical chopper modulating the excitation beam at 2 kHz.

The use of engineered single-crystal diamond leads to high quality surfaces, which, in turn, allows us to implement optical and spin characterization protocols otherwise difficult to replicate. We selectively probe the fluorescence stemming from individual NV layers in the stack with our microscope configured in "confocal" mode upon a simple reconfiguration of the excitation beam path (Fig. 1a). For illustration purposes, Fig. 1d shows representative photoluminescence (PL) spectra from a ~1 μm² section in one of the membranes for different laser powers. A decomposition into negative and neutral contributions yields an intensity-dependent NV⁻ fractional population evolving from ~30% at the lowest laser powers to saturate at ~60%, below the 75% limit characteristic of bulk NVs under 532 nm illumination[27]. This lower-than-normal NV$^-$ fraction is indicative of partial NV$^-$ ionization in the dark, likely due to tunneling of the excess electron to adjacent traps[34,35] and/or capture of itinerant near-surface holes[36].

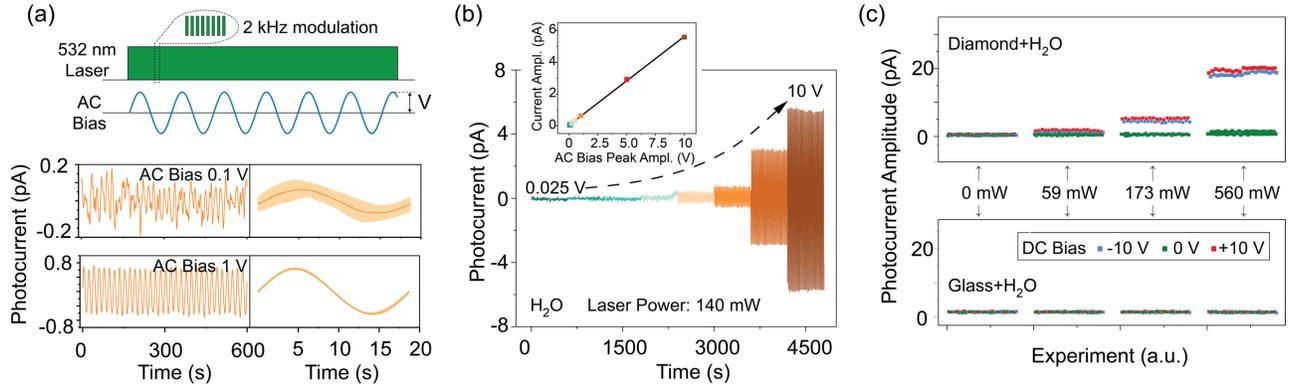

**Fig. 2: Detection of photocurrent in water.** (a) Experimental protocol (top) and example photo-current signals (bottom left) for pure $H_2O$; the AC bias frequency is 53 mHz, the lock-in integration constant is 1 s, and the laser power is 140 mW. The lower right plots are one-cycle averages, here used for error determination. (b) Observed photo-current temporal signals from $H_2O$ for varying AC voltage bias; the 532 nm laser power is 140 mW. The upper insert shows the photo-current peak amplitude vs the AC bias peak amplitude as derived from the main plot. (c) $H_2O$ photocurrent signal amplitudes with and without applied DC bias for varying laser intensities when in contact with a diamond or a glass stack of membranes (left and right plots, respectively). In all experiments, the beam illuminates a 2-mm-diameter spot extending uniformly across all membranes.

Observations of optically detected magnetic resonance (ODMR) — possible in NV$^-$ thanks to its spin-dependent photon emission[37,38] — provide additional clues on the surface characteristics. The upper plot in Fig. 1e shows the ODMR spectrum obtained upon a microwave (MW) sweep across the NV$^-$ zero-field resonance[39]: We attain an optical spin contrast of ~6%, lower than the 30% optimum but better than typically observed in similarly shallow NV ensembles[40]. The measured linewidth (~8 MHz) is moderate, especially if we consider the underlying $^{14}N$ hyperfine broadening and the characteristically large strain- and electric-field-induced heterogeneity of the NV$^-$ crystal field near the surface. On the other hand, the relatively short spin-echo lifetime (~2 μs, not shown) is characteristic of NVs subjected to magnetic and electric noise near the surface[41-42,43].

We interrogate coexisting paramagnetic centers via NV-detected double electron-electron resonance (DEER) experiments (low insert in Fig. 1e). The DEER spectrum features two dips of comparable amplitude, indicative of multiple classes of spin-active defects[44-46]. Interestingly, we find no characteristic PL dips at the hyperfine-shifted frequencies of neutral substitutional nitrogen[46] ($N_S^0$) suggesting the loss of the donor electron and thus the formation of $N_S^+$, a spin-less charge state. Likely the result of incomplete surface oxidation[8], we interpret these observations as indicative of a positive band bending strong enough to deplete $N_S^0$ (a shallower donor) but only partially affecting NV$^-$. We return to these important considerations in the following section.

**Photocurrent measurements in water**

Figure 2a introduces our experimental protocol comprising laser excitation and an alternating (AC) voltage bias between the electrodes, here introduced to minimize space charge fields[47] (see Methods). Because the AC frequency (in the mHz range) is much slower than the laser modulation rate (2 kHz) and lock-in inverse integration time (typically, 1 s$^{-1}$), each point in the photocurrent plot can be seen as an "instantaneous" measurement under a varying direct (DC) bias field. The lower plots in Fig. 2a display representative data sets under 532 nm excitation and AC bias of different maximum amplitude; using one-cycle averages to quantify the error (bottom right plots), we benchmark the photocurrent sensitivity of our setup at about 0.6 pA Hz$^{-1/2}$.

The use of a sinusoidal shape to fit our observations necessarily hinges on a linear relation between the photocurrent and applied AC bias, a condition we verify via measurements under varying AC field amplitudes (Figs. 2b); the proportionality we observe mirrors the Ohmic response we find in our water-filled chip in the absence of optical excitation. Importantly, we measure zero photocurrent — even at the highest possible laser intensities — if we replace the membranes by an equal number of glass slides arranged in identical geometry, which simultaneously demonstrates the key role of the diamond–water interface as well as our ability to separate the photo-generated current from background (i.e., light-insensitive) contributions (Fig. 2c).

Although the system at hand has been explicitly designed to make NVs abundant, the nature of the point defect serving as the carrier source is difficult to disambiguate as ion implantation leads to concomitant point defects potentially susceptible to charge state changes under optical illumination. An initial route to deconvolving contributions from NVs and these coexisting sources involves the characterization of photocurrent under illumination of variable wavelength. We capture these experiments in Fig. 3a: In the absence of a light source with the required tunability and output power, we adapt the setup in Fig. 1a to accommodate continuous wave (cw) lasers at discrete wavelengths throughout the visible range. Relative to our observations at 532 nm (Fig. 2), we find a quick jump of the photocurrent amplitude at shorter wavelengths, with a ~30-fold increase at 450 nm; as the wavelength grows, however, we measure a monotonic decrease to attain virtually no response at 632 nm (and beyond).

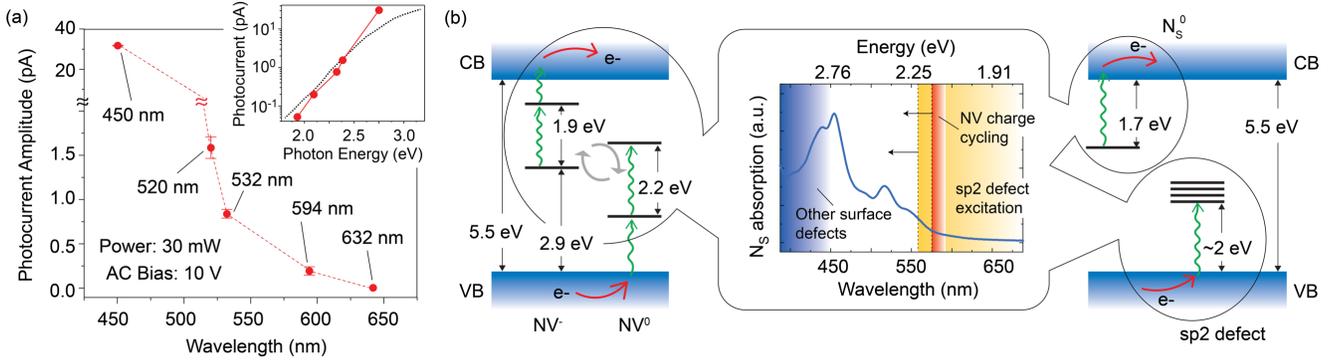

**Fig. 3: Photocurrent response at varying wavelength.** (a) Photocurrent amplitude as a function of the excitation wavelength. The inset shows the same data set but in logarithmic scale vs photon energy; the dashed black line reproduces the data set in Ref. [49]. In all cases, the AC voltage bias is 10 V and the excitation laser power is 30 mW for an illuminated area of ~2 mm$^2$; all other conditions as in Fig. 2. (b) (Main plot) Calculated $N_S^0$ ionization cross section vs wavelength as derived from density functional theory (solid trace). The red band indicates the range where two-photon NV charge cycling activates, see Ref. [27]; orange and blue areas respectively indicate activation bands of single-photon electron injection into unoccupied states in primal sp2 defects and other surface defects as derived from Refs. [52] and [8], respectively. (Left insert) Schematics of NV two-photon ionization and recombination, left- and right-hand side diagrams, respectively. (Right insert) Single photon processes corresponding to $N_S^0$ photoionization and electron injection into surface traps (top left and bottom right diagrams, respectively). CB: Conduction band; VB: Valence band.

While the above findings are insufficient to expose a specific point defect as the dominant source, the observed spectral dependence does contain some important clues. For example, the trend at longer wavelengths (i.e., 594 and 632 nm) is not inconsistent with that expected for NVs because the two-step, one-photon processes driving NV charge cycling at 532 and 520 nm get suppressed above 575 nm (where the photon energy is insufficient to excite NV$^0$). Phonon-assisted recombination at longer wavelengths makes this transition gradual[27,48], hence implying an NV-dominated photocurrent would arguably fall off smoothly, as observed. As a matter of fact, this wavelength dependence qualitatively reproduces that observed for NV ensembles in bulk diamond[49] (insert in Fig. 3a).

Unfortunately, this trend is not unique and hence insufficient to separate NV centers from other potential charge sources. Among them, substitutional nitrogen ($N_S$) is a natural candidate: Despite its 1.7 eV (i.e., 730 nm) ionization threshold, large structural reconfigurations[50,51] required for $N_S^0 \rightarrow N_S^+$ conversion make optical excitation inefficient for photon energies below ~2.1 eV, hence pushing free carrier activation to wavelengths below ~590 nm. Since the observed photo-current should reflect the $N_S^0$ absorption cross section — explicitly calculated in Fig. 3b via density functional theory (DFT), see Methods — one would also expect in this case a gradual signal change in the 500 – 650 nm range, not too different from our experimental results.

Assigning the photocurrent spectral response to $N_S^0 \rightarrow N_S^+$ conversion, however, seems unwarranted, particularly given the absence of $N_S^0$ DEER signal discussed above. Further, the photo-current response at 450 nm is about 30-fold larger than that observed at 532 nm, and hence disproportionally big for a process solely dominated by nitrogen defects (see Fig. 3b as a reference). The origin of this steep change at shorter wavelengths is presently unclear, but we hypothesize it stems from charge injection produced by excitation of unoccupied surface states[8,22]. These take varying forms depending on the fluid in contact with the crystal and surface termination protocol, but they are known to activate under blue excitation and are generally abundant in chemically oxidized diamond surfaces[8]. Similarly, the response above ~520 nm could be due to primal sp$^2$ surface defects, recently shown to serve as electron traps in the 1.5 – 2.2 eV range above the valence band maximum[52]. Upward band bending takes place near the surface as these electronic traps (partially) fill out in nitrogen-doped diamond[53], thus explaining the depletion of $N_S^0$ and the lower-than-normal concentration of NV$^-$. Further, electron capture renders these defects paramagnetic, thus providing a rationale for the observed

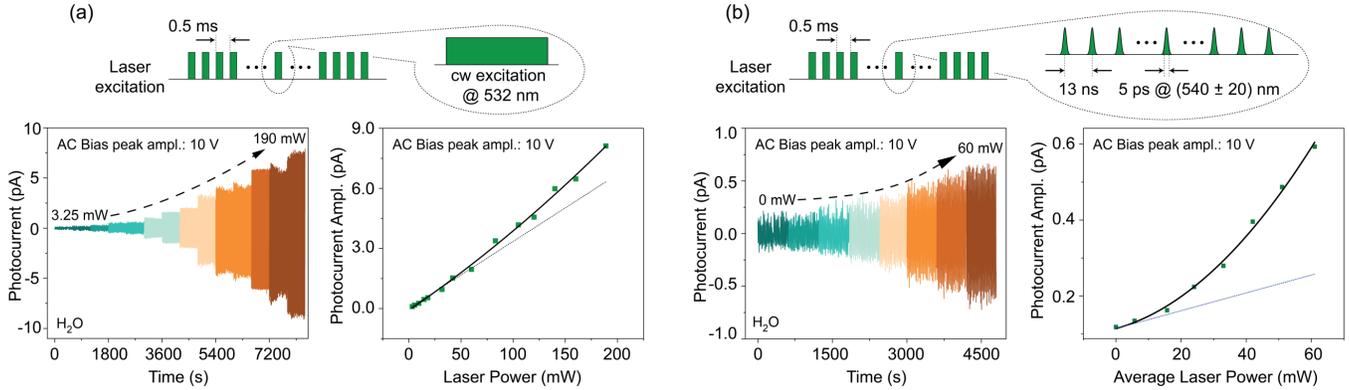

**Fig. 4: Photocurrent response at varying laser power.** (a) We reproduce the protocol in Fig. 2a, except that this time we gradually change the 532 nm laser power for a fixed AC bias amplitude (10 V). The lower right plot shows the photocurrent amplitude as extracted from the raw data (left plot). The solid line is a quadratic fit with a linear component plotted separately as a dashed line. (b) Same as in (a) but for light excitation from a supercontinuum laser producing a 78 MHz train of 5-ps-long pulses with a wavelength dispersion of 40 nm centered at 540 nm. Unless otherwise stated, working conditions are those in Fig. 2.

DEER signal; note that in the absence of spin-active nuclei — other than 1%-abundant $^{13}C$ — magnetic resonance signals should not display hyperfine satellites, consistent with our observations in Fig. 1e.

In a scenario where NVs coexist with surface traps, one route to separating individual photo-current contributions is to exploit their distinct charge state responses to light, respectively governed by two- or single-photon processes, and thus leading to quadratic or linear signal changes under varying illumination intensity. Figure 4a extends the observations in Fig. 2 to examine the photocurrent signal for variable laser powers. We limit these experiments to 532 nm excitation, chosen to ensure significant NV charge cycling while restricting the type of surface defects to primal $sp^2$ traps (featuring unoccupied levels accessible to this wavelength[52]). Analogous to photocurrent measurements in bulk Type 1b diamond — often dominated by single photon charge injection processes[54] — the response we observe is predominantly linear, even though a quadratic correction is clearly necessary to attain a good fit. Given the one-photon nature of electronic injection in $sp^2$ defects, our results point to these surface traps, not the NVs, as the dominant carrier source in these experiments. Interestingly, photoelectrically detected magnetic resonance[54,55] (PDMR) experiments — implemented herein through the combined use of our optical excitation and MW capabilities — yields no observable signal at the characteristic $NV^-$ crystal field frequency, which indirectly supports the above conclusion.

The small but discernible quadratic contribution we extract from Fig. 4a, however, suggests NV centers can indeed play a significant role if the illumination intensity is sufficiently high. We validate this idea in Fig. 4b where, rather than cw excitation, we make use of a supercontinuum laser producing a train of picosecond-long pulses at a repetition rate of 78 MHz (see Methods). To bring the average power to a range comparable to that attained with the cw light source, we set a 40 nm spectral window centered at 540 nm. Remarkably, we find that increasing the (average) laser power leads to a predominantly parabolic signal growth, indicative of NV-dominated photocurrent. This change can be rationalized as a manifestation of the quadratic power dependence inherent to two-photon ionization/recombination processes, and hence the augmented impact of picosecond pulses on the NV charge-cycling dynamics. As an important side note, attempts to observe a PDMR signal from NVs exposed to pulsed illumination yielded inconclusive results. In this case, we attribute the lack of a clear spectral signature to the combination of a weak photocurrent signal (in the fA range under present conditions), and the rather modest PDMR contrast (typically 3–10% in ensembles). This limitation could be circumvented in future experiments through the use of higher power femtosecond lasers operating in the visible range.

**Conclusions and outlook**

In summary, we leveraged diamond engineering and microfluidics to investigate photo-activated carrier transfer into water. Unlike most prior work — typically relying on boron-doped, hydrogen-terminated diamond — here we focused on oxygen-terminated, electronic grade single crystals engineered to host near-surface NV centers; in these samples, chemical oxidation counters band bending, and hence maintains a significant fraction of NVs in the negatively charged state. We detect photocurrent throughout the visible range, although the amplitude decreases with longer wavelengths to become negligible above 600 nm. Combined with ODMR and photocurrent measurements under varying laser power and wavelength, our observations expose photocurrent contributions from NV centers and other co-existing point defects, most likely associated with surface acceptors. This latter group — probably formed by primal $sp^2$ defects or dangling bonds (or both) — dominates photocurrent injection under weaker green illumination; pulsed excitation, on the other hand, efficiently activates two-photon processes, ultimately making the NV contribution prevalent.

Future work should identify the conditions for optimal carrier injection without compromising, however, on the concentration of NV$^-$. Similarly, rendering NVs the main source of photogenerated carriers will rely on new strategies able to improve the NV formation efficiency during the anneal[56] and/or the use of pulsed excitation with time-averaged power higher than that possible here. Additional theoretical and experimental work will also be necessary to better understand the mechanisms responsible for replenishing carriers in the color center ensemble, mitigate the formation of space charge fields, best assess the role of heterogeneity in surface termination, and determine fractional contributions to the photocurrent from injected electrons and holes.

The use of single-crystalline diamond hosting shallow, long-spin-coherence paramagnetic centers opens intriguing opportunities for the investigation of solvated-carrier-initiated chemical reactions. Of particular interest are experiments in nanoscale confined media — produced, e.g., with the help of patterned two-dimensional materials[57] — where the proximity to shallow NV$^-$ could be simultaneously exploited to implement optical-magnetic-resonance-based sensing approaches[58]. These ideas could prove especially fruitful in the investigation of ionic transport in nanofluidic channels, and their application to fluidic neuromorphic computing[12]. While the focus here centered on liquids, electron injection into vacuum seems similarly possible[23], and could be leveraged to develop new forms of single-NV-based electron microscopies, an interesting feature being the ability to monitor the number of photogenerated carriers via, e.g., single shot NV charge state readout[59,60].

Another attractive possibility entails the implementation of alternate dynamic nuclear polarization strategies via direct injection of spin-polarized carriers into the fluid or through their capture by surface defects in direct interfacial contact. In the case of the NV, the generation of spin polarized electrons could be accomplished by proper optical initialization preceding ionization[61]; similarly, double resonance schemes could be adapted to transfer spin polarization to surface paramagnetic centers[62] followed by targeted optical ionization. Mirroring hyperpolarization techniques such as SPINOE[63], injection of spin polarized carriers may open new pathways for the transfer of polarization to a fluid[13], thus far inefficient due to losses introduced by surface paramagnetic impurities during spin diffusion[14] (a comparatively slow process). Combined with photocurrent measurements, this class of experiments should also prove relevant in the investigation of spin-dependent charge transfer across solid-liquid interfaces, demonstrated recently though requiring strong spin-polarizing fields[15].

**Methods**

***Sample characteristics***: We utilize fourteen customized [100] electronic-grade diamond membranes provided by Qnami AG. Each membrane has dimensions 3×3×0.05 mm$^3$ and features a 6-μm-deep, 2.5-mm-wide, 3-mm-long trench produced via reactive ion etching on one of its surfaces. For a stacked set of membranes, these trenches create 6-μm-wide openings effectively serving as channels for liquid flow (see Fig.1). $^{14}$N ion implantation with energy of 2.5 keV and a dose of $1\times10^{13}$ ions/cm$^2$ leads to a 5-nm-deep layer of nitrogen spread over a ~2 nm range as derived from SRIM simulations[64]. To induce NV center formation, known annealing protocols were used[65], except that the end temperature was 1000 C instead of 800 C. Following annealing, a tri-acid mixture was used for cleaning and oxygen-terminating the diamond surface.

***Microfluidic device***: The membranes are placed within a custom-built 3D-printed microfluidic cell, and sealed with a thin glass layer so as to enable laser excitation. The microfluidic cell is equipped with two ports for the flow of aqueous solutions past the diamond membrane's channels and a pair of planar, gold-coated electrodes proximal to the membrane's edge. These electrodes connect to a voltage source and a lock-in amplifier operating as an ammeter (Fig. 1c). An automated syringe pump gives us precise control over the liquid flow rate, which we keep at the minimum required to mitigate the formation of space charge fields.

***Experimental setup***: Our system can be alternatively configured for photocurrent experiments or for optical characterization of the NV ensemble. In photocurrent measurement mode, we illuminate the diamond stack with a 1-W laser collimated over a ~2×2 mm$^2$ area, which remains approximately uniform throughout the membrane stack. The photocurrent from the electrode is measured by a lock-in amplifier, with the current signal locked to the chopper frequency. A set of cw diode lasers, each operating at a different color allows us to investigate the photocurrent response at different illumination wavelengths; for pulsed excitation, we resort to a supercontinuum laser (NKT) whose wavelength range we control with a set of integrated filters; the pulsed duration is 5 ps and the repetition rate is 78 MHz.

A flip mirror switches the system to a confocal microscope. In this mode, we choose a target membrane from the stack and steer the photoluminescence to an avalanche photodetector or to a spectrometer, depending on the end application; in this operating mode, we collect fluorescence from a ~5-μm-long focal volume, and the illuminated area is ~1 µm$^2$. Continuous-wave ODMR under aqueous conditions uses the MW generated by an omega-shaped antenna patterned on a PCB board. We use a pulse controller as well as optical and MW switches to program arbitrary time-resolved ODMR protocols including the Hahn-echo and DEER sequences.

*Magnetic resonance:* We characterize surface NVs and coexisting paramagnetic centers with the help of optically detected magnetic resonance techniques. Continuous wave ODMR (upper plot in Fig. 1e) records the NV fluorescence resulting from 532 nm, 0.6 mW laser illumination in the presence of cw MW excitation of variable frequency (respectively, green and blue blocks in the figure insert). The DEER protocol (lower plot in Fig. 1e) comprises initialization and detection laser pulses (532 nm, 0.6 mW, green blocks) enclosing a fixed-duration Hahn echo sequence resonant with the NV $^3A_2$ $|m_S = 0\rangle \leftrightarrow |m_S = -1\rangle$ transition (blue blocks in the insert). The DEER spectrum emerges as one varies the frequency of a second MW π-pulse (nearly) overlapping in time with the Hahn-echo inversion pulse[46] (red block in the insert).

*Density functional theory.* To calculate the $N_S^0$ absorption cross section we use the PAW method[66] with Perdew-Burke-Ernzerhoff (PBE)[67] and range-separated Heyd-Scuseria-Ernzerhoff (HSE06)[68] functionals to account for electronic exchange-correlation interactions during atomic relaxations and self-consistent (SCF) calculations, respectively. For the plane-wave basis, we use a cut-off of 370 eV yielding well-converged results[49,69,70]. We attain equilibrium defect structures by embedding the nitrogen impurity in a 4×4×4 (512-atom) pristine diamond supercell and relaxing the atoms until net forces are below 10$^{-3}$ eV/Å. During SCF calculations, electronic loops converge to energy differences below 10$^{-8}$ eV. We sample the Brillouin zone only at the Γ-point in all supercell calculations. One exception, though, is when calculating the $N_S^0$ wave function derivatives (used for obtaining transition dipole moments in the momentum representation), in which case we employ a 6×6×6 *k*-point Γ-centered sampling mesh, enough to attain good convergence. We obtain the ionization cross section following the scheme described in Ref. [49]. In particular, we correct the $N_S^0$ ionization response with the Frank-Condon shift to account for the large atomic reconfiguration at $N_S^0$ electronic ionization threshold[50,51].

**Data availability**

The data that support the findings of this study are available from the corresponding author upon reasonable request.

**Acknowledgments**

We thank Alexander Pines for early discussions and help procuring the diamond membranes. We are indebted to Artur Lozovoi for valuable assistance with the experimental setup. We also acknowledge Nicolas Giovambattista, Gustavo Lopez, Alastair Stacey, Jonathan Owen, and Nathalie de Leon for helpful insight. K.X. and C.A.M acknowledge support from the National Science Foundation through NSF-2223461; D.P. acknowledges funding from the National Science Foundation via grant NSF-2203904. A.W. acknowledges funding from the National Science Foundation via grant NSF-2112550. G.I.L.M. acknowledges support from the National Science Foundation via fellowship NSF-2208863. All authors acknowledge access to the facilities and research infrastructure of the NSF CREST IDEALS, grant number NSF-2112550. The Flatiron Institute is a division of the Simons Foundation.

**Author contributions**

K.X., D.P., A.W. and C.A.M. conceived the experiments. K.X., A.W., D.P. conducted the experiments, G.I.L.M and J.F. carried out the DFT calculations. C.A.M. supervised the project and wrote the manuscript with input from all authors.

**Competing interests**

The authors declare no competing interests.

**Correspondence**

Correspondence and requests for materials should be addressed to C.A.M.